\documentclass[preprint,12pt]{elsarticle}

\usepackage{amssymb}

\begin{document}

\begin{frontmatter}



\title{Off-diagonal geometric phase of atom-electron coupling in hydrogen atom}


\author{Guo-Qiang Zhu}

\address{Laboratory for Quantum Information and Department of Applied Physics, China Jiliang University, Hangzhou 310018, P.R. China}

\begin{abstract}
In this paper, the off-diagonal geometric phase of thermal state in
hydrogen atom under the effects of external magnetic field is
considered. Increasing temperature tends to suppress the
off-diagonal geometric phase, including $1$-order and $2$-order
cases. On the other hand, the relationship between the geometric
phase and external magnetic field is discussed.
\end{abstract}


\end{frontmatter}


\section{Introduction}\label{sect:introduction}
The concept of geometric phase (GP) was first introduced by
Panchartnam in his study of interference of classical light in
distinct states of polarization \cite{pancharatnam}. Berry's work
showed a quantum pure state can retain the information of its motion
when it undergoes a cyclic evolution \cite{berry}. Simon
\cite{simon} subsequently recasted the mathematical formation of
Berry phase with the language of differential geometry and fibre
bundles. He observed that the origin of Berry phase is attributed to
the holonomy in the parameter space. Due to its robustness to
imperfections, such as docoherence and the random unitary
perturbations, geometric phase has many applications in the field of
quantum information processing and condensed matter physics. It has
been pointed out that the non-Abelian holonomy may be used in the
construction of universal sets of quantum gates for the purpose of
achieving fault-tolerant quantum computation
\cite{zanardi1,zanardi2}. On the other hand, GP, being a measure of
the curvature of the Hilbert space, when associated with the energy
crossing has a peculiar behavior near the degeneracy point. GP may
be considered as a good candidate for a universal order parameter
for quantum phase transitions \cite{carollo,slzhu}.

In the early discussions, most researches were focused on the
evolution of pure states. In realistic world, due to the effect of
environment, the most states are mixed. Uhlmann was probably the
first to address the issue of mixed state holonomy, but as a purely
mathematical problem \cite{uhlmann1,uhlmann2}. Later Sj\"{o}qvist
\emph{et al.} discussed the geometric phase for non-degenerate mixed
state under unitary evolution in Ref.\cite{sjoqvist}, basing on the
Mach-Zender interferometer.  Singh \emph{et al.}\cite{singh} gave a
kinematic description of the mixed-state GP in Ref.\cite{sjoqvist}
and extended it to degenerate density operators. The generalization
to nonunitary evolution has also been adressed in Refs.\cite{tong1}.

Later it was  found that the above diagonal GP itself could not
exhaust al information contained in phases acquired when the quantal
system undergoes an adiabatic evolution. The notions of GP break
down in cases where the inteference visibility between the initial
and final states vanishes. This gives rise to the definition of the
off-diagonal geometric phase factor \cite{Manini}. Here, The concept
of Berry phase was extended to the the evolution of more than one
state.  They found a set of independent off-diagonal phase factors
that exhaust the geometrical phase information carried by the basis
of eigenstates along the path. Later, the concept of off-diagonal GP
was extended to the cases of mixed states\cite{filipp}.

The off-diagonal geometric phase factor of the degenerate mixed
states was defined as \cite{tong}
\begin{equation}\label{off-phase}
\gamma^{(l)}=\Phi\left[Tr(\sum_{\alpha=1}^l
U(\tau)V_{j_{\alpha}}^{\parallel}(\tau)\sqrt[l]{\rho_{j_{\alpha}}})\right]
\end{equation}
where $\Phi[z]\equiv z/|z|$ for any nonzero complex number $z$. The
phase is manifestly gauge invariant. $U(\tau)$ is the unitary
evolution operator and $U(\tau)=\exp(-i\tau H/\hbar)$ and $H$ is the
total Hamiltonian. Throughout this paper, Planck's constant $\hbar$
is set to unity for simplicity. One can define a set of
noninterfering density operators,
\begin{equation}
\rho_n=\lambda_1 P_{n;1}^{(m_1)}+\cdots\lambda_{K}P_{n;K}^{(m_K)},
n=1,\ldots,N,
\end{equation}
where $P_{n;k}^{(m_k)}=W^{n-1}P_{1;k}^{(m_k)}(W^{\dagger})^{n-1}$
and $W$ is a permutation unitarity as
$W=|\psi_1\rangle\langle\psi_N|+|\psi_N\rangle\langle\psi_{N-1}|+\cdots|\psi_2\rangle\langle\psi_1|$.

 The parallel transport unitary operator for $\rho_n$ may
be expressed as $U_n^{\parallel}(t)=U(t)V_n^{\parallel}(t)$ with
supplementary operators
$V_n^{\parallel}(t)=\alpha_{n;1}^{\parallel}(t)+\dots+\alpha_{n;k}^{\parallel}(t)$,
where
\begin{equation}
\alpha_{n;k}^{\parallel}=P_{n;k}^{m_k}\mathbf{T}\exp\left(-\int_0^t
P_{n;k}^{m_k}U^{\dagger}(t')\dot{U}(t')P_{n;k}^{m_k}dt'\right)P_{n;k}^{m_k}.
\end{equation}
$\mathbf{T}$ denotes time ordering. $P_{n;k}^{m_k}$ is the projector
of rank $m_k$ onto the $m_k$-fold degenerate eigenspace.

When the Hamiltonian is independent of the time $t$,
$\alpha_{n;k}^{\parallel}$ will be reduced as follows,
\begin{equation}
\alpha_{n;k}^{\parallel}=P_{n;k}^{m_k}
\mathrm{T}\exp(itP_{n;k}^{m_k} H P_{n;k}^{m_k})P_{n;k}^{m_k}.
\end{equation}
Apparently $\gamma^{(1)}$ are the standard mixed-state geometric
phase factors associated with the unitary paths in state space. The
off-diagonal mixed-state geometric phases contain information about
the geometry of state space along the path connecting pairs of
density operators, when the standard mixed-state geometric phases
are undefined. The uncontained information can be shown via
high-order off-diagonal geometric phase. The $l=2$ case has been
discussed in terms of two-particle interferometry \cite{l=2}.

The off-diagonal geometric phase for some models has been discussed.
For example, X.X. Yi \emph{et al.}\cite{intersystem} investigated
the effect of the intersubsystem coupling on the off-diagonal
geometric phase in a composite system, where the system undergoes an
adiabatic evolution.

\section{Model}
In this paper, the thermal state of the hydrogen atom is discussed.
As one knows, in the hydrogen atom, the electron spin is coupled to
the nuclear spin by the hyperfine interaction. The hyperfine line
for the hydrogen atom has a measured magnitude of 1420 MHz in
frequency. Some calculation on the basis of first-order perturbation
for the magnetic dipole interaction between the electron and nucleus
gives contribution to the coupling strength of $\mathbf{I}\cdot
\mathbf{S}$ term. Taking account of the effect of an external
magnetic field, one can have a model Hamiltonian which is \cite{bec}
\begin{equation}\label{h0}
H_0=J(I_x\otimes S_x+I_y\otimes S_y+I_z\otimes S_z)
\end{equation}
where $J$ is the coupling constant. Here the electronic orbital
angular momentum $L$ is assumed to be zero.

As we know, for a hydrogen atom, the nucleus and electron has both
spin-1/2, $\mathbf{I}=\mathbf{S}=\frac{1}{2}\mathbf{\sigma}$.  The
Hamiltonian Eq.\ref{h0} has $2$ district eigen-energies: $1/4J$ and
$-3J/4$ and the former is 3-fold degenerate with eigenvestors:
$|\phi_i\rangle,i=1,2,3$ and the latter is non-degenerate with
eigenvector $|\phi_4\rangle$. The four eigenvectors are given by
\begin{eqnarray*}
|\phi_1\rangle&=&|00\rangle\\
|\phi_2\rangle&=&\frac{1}{\sqrt{2}}(|01\rangle+|10\rangle)\\
|\phi_3\rangle&=&|11\rangle\\
|\phi_4\rangle&=&\frac{1}{\sqrt{2}}(|01\rangle-|10\rangle)
\end{eqnarray*}

Now the atom is assumed to be under the temperature $T$. Due to the
effect of the temperature, the state of the atom is highly mixed.
The density matrix can be written in terms of partition function $
\rho=e^{-\beta H}/tr e^{-\beta H},$ where $\beta=1/kT$ and $k$ is
Boltzmann's constant and it is set to unity throughout this paper.
$T$ denotes the temperature. Due to the effect of thermal
fluctuation, the system is highly mixed. At the initial time $t=0$,
no magnetic field is imposed so the density matrix of the initial
state is given by
\begin{equation}
\rho_0=\frac{e^{-\beta H_0}}{tr e^{-\beta H_0}}.
\end{equation}

One can see the above $4$ eigenvectors are also the eigenvectors of
$\rho_0$. After the initial time $t=0$, the external magnetic field
is imposed upon the system and the Hamiltonian will be $H=H_0+H_I$.
The degeneracy of the state is destroyed due to the magnetic field.
The interaction term $H_I$ is given by
\begin{equation}\label{interaction}
H_I=C I\otimes S_z+D I_z\otimes I.
\end{equation}
The parameters $C$ and $D$ are related with external magnetic
fields,
$$C=g\mu_B B, \ \ D=-\frac{\mu}{I}B.$$  The nuclear magnetic moment $\mu$
equals $2.793\mu_N$ where $\mu_N=e\hbar/(2m_p)$. In general C is
much larger than D, $|C/D|\sim m_p/m_e\approx 2000$, so for most
applications D may be neglected. At the same level of approximation
the $g$ factor of the electron may be put equal to 2. Therefore one
has $$H_I=g\mu_B B I\otimes S_z.$$ It is obvious that the
commutation $[H_0,H_I]\neq0$. It implies that the external magnetic
field  will drive the system to involve with the time. At time $t$,
the state is
$$
\rho(t)=U(t)\rho_0 U^{\dagger}(t)
$$
where the unitary matrix $U(t)=\exp\{-i H t\}$.  Assuming after a
period $T$, the state evolves to its initial state $\rho(0)$, i.e.,
$\rho(0)=\rho(T)$, then one has
\begin{equation}
T=\frac{2n\pi}{\sqrt{C^2+J^2}}.
\end{equation}
Here we restrict ourselves to  $n=1$. Here $C=g\mu_B B$ which is the
function of magnetic field $B$, so one can see the magnetic field
$B$ and coupling constant $J$ can control the period $T$. After a
period evolution, a mixed-state geometric phase can be observed. In
next section, we will discuss this mixed-state geometric phase.

\section{Geometric Phase}
It is obvious that the initial mixed state $\rho_0=e^{-\beta H}/tr
e^{-\beta H}$ is degenerate. In Ref.\cite{temperature}, A.T.
Rezakhani and P. Zanardi studied the GP of an open quantum system
interaction with a thermal environment by using the definition given
in Ref.\cite{tong}. In sect.\ref{sect:introduction}, the shortcoming
of the definition has been discussed. To overcome the shortcoming,
we will use the definition of off-diagonal GP given by
Eq.\ref{off-phase} to evaluate the GP of the state to study the
information which are not exhausted by diagonal GP.

\subsection{1-order GP}
The $1$-order off-diagonal GP is equivalent to the ordinary diagonal
GP which is thought not to exhaust the information about the
geometry of state along the path connecting pairs of density
operators.

The initial density matrix $\rho_1$ can be written as
$\rho_1=\sum_{i=1}^3
\lambda_1|\phi_i\rangle\langle\phi_i|+\lambda_2|\phi_4\rangle\langle\phi_4|.
$ There are two eigenspaces. The first space is $3$-Dimensional and
the second one is $1$-dimensional. From Eq.\ref{off-phase}, the
$1$-order off-diagonal GP can be reduced to
\begin{equation}
\gamma^{(1)}=\Phi\left(tr U(\tau)V_1^{\parallel}\rho_1\right),
\end{equation}
where $U(\tau)$ is the unitary operator.
$V_1^{\parallel}=\alpha_{1;1}^{\parallel}(t)+\alpha_{1;2}^{\parallel}(t)$
which are the unitary on the first and second degenerate subspaces.
\begin{figure}
\begin{center}
\includegraphics[width=8cm]{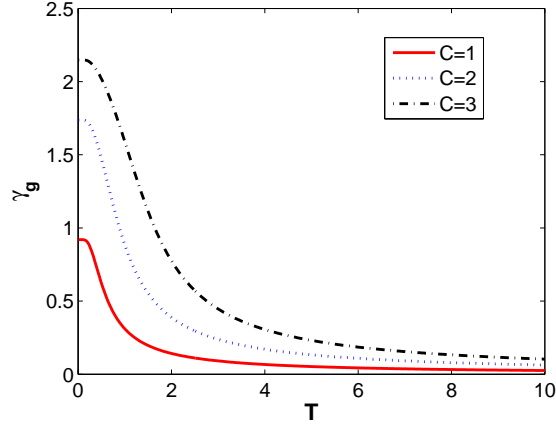}
\caption{\label{1order:T}First order off-diagonal GP $\gamma_g$
versus temperature $T$, when $C=g\mu_B B$ are 1,2,3 respectively.}
\end{center}
\end{figure}

\begin{figure}
\begin{center}
\includegraphics[width=8cm]{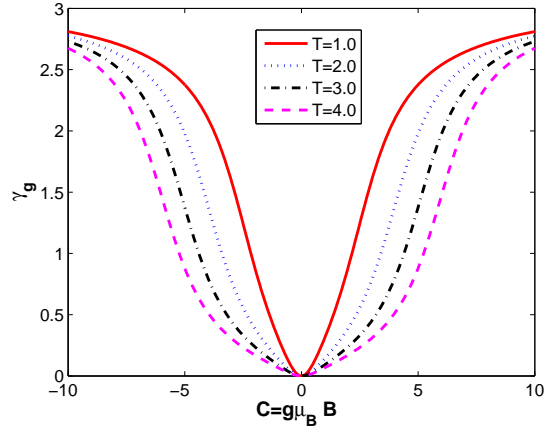}
\caption{\label{1order:C}First order off-diagonal GP $\gamma_g$
versus external magnetic field $C=g\mu_B B$, when the temperature
$T=1.0,2.0,3.0,4.0 $ respectively.}
\end{center}
\end{figure}

Fig.\ref{1order:T} shows the effect of temperature upon GP, one can
see increasing temperature tends to suppress GP. Fig.\ref{1order:C}
shows the relation between GP and the external magnetic field
$B=C/g\mu_B$. One can see when $B$ changes its direction, GP is not
changed. When increasing of  the length of magnetic field tends to
enhance GP. So that one can control GP via manipulating the magnetic
field. From the figure, one can see that for a relatively large
temperature interval geometric phase is varying very slowly with
temperature. When $C=0$, the geometric phase vanishes. It is trivial
because in this case no magnetic field is imposed so that the state
will not evolve with the time. When the temperature approaches
infinity, i.e., $\beta$ approaches zero, the matrix $e^{-\beta H}$
will become an identity matrix, so that the unitary matrix $U(t)$
will be commutative with the initial state and the state will not
evolve with the time. There is no GP can be detected.

\subsection{2-order GP}
\begin{figure}
\begin{center}
\includegraphics[width=8cm]{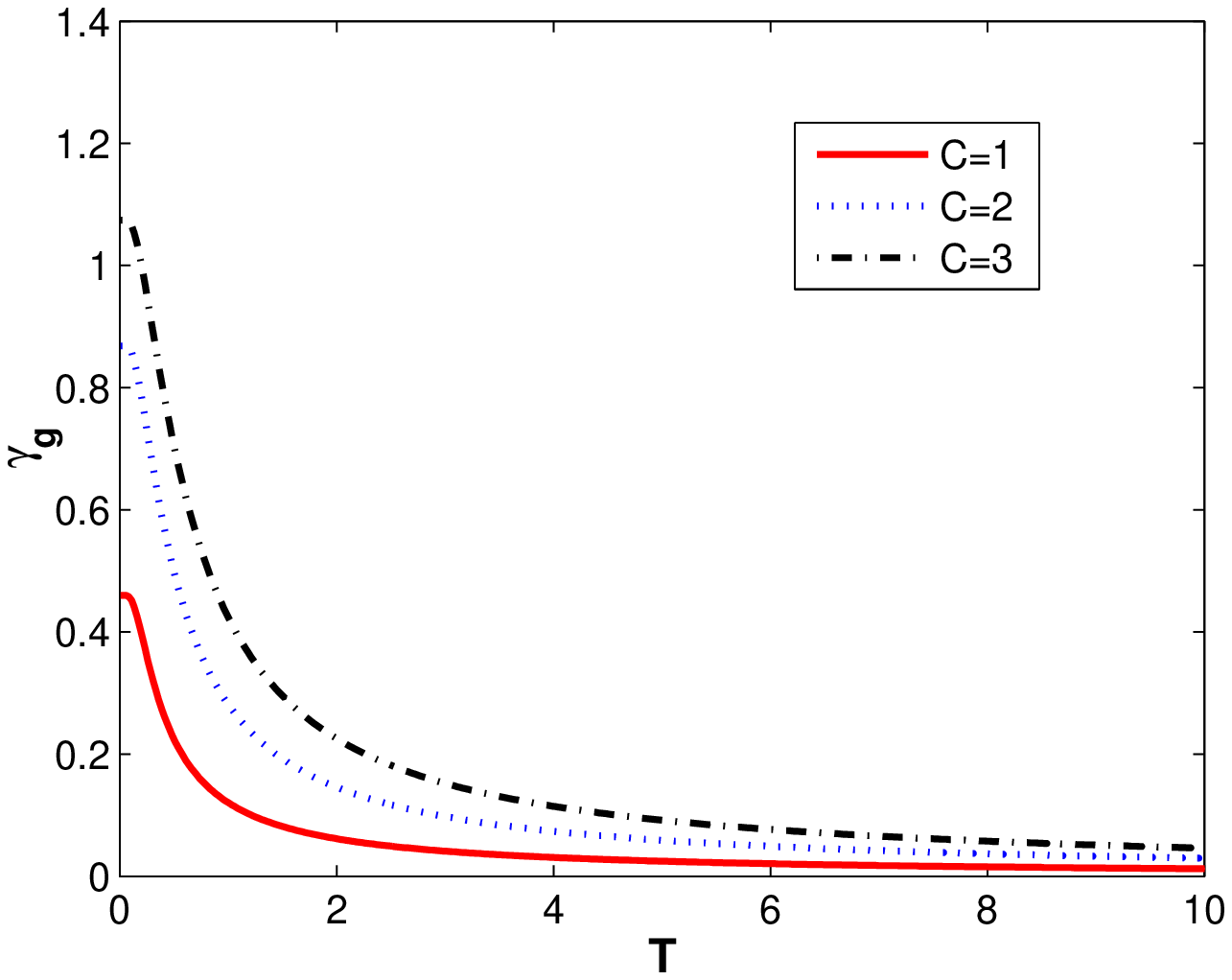}
\caption{\label{2order:T}Second order off-diagonal GP$\gamma_g$
versus temperature $T$, when $C=g\mu_B B$ are 1,2,3 respectively.}
\end{center}
\end{figure}
\begin{figure}
\begin{center}
\includegraphics[width=8cm]{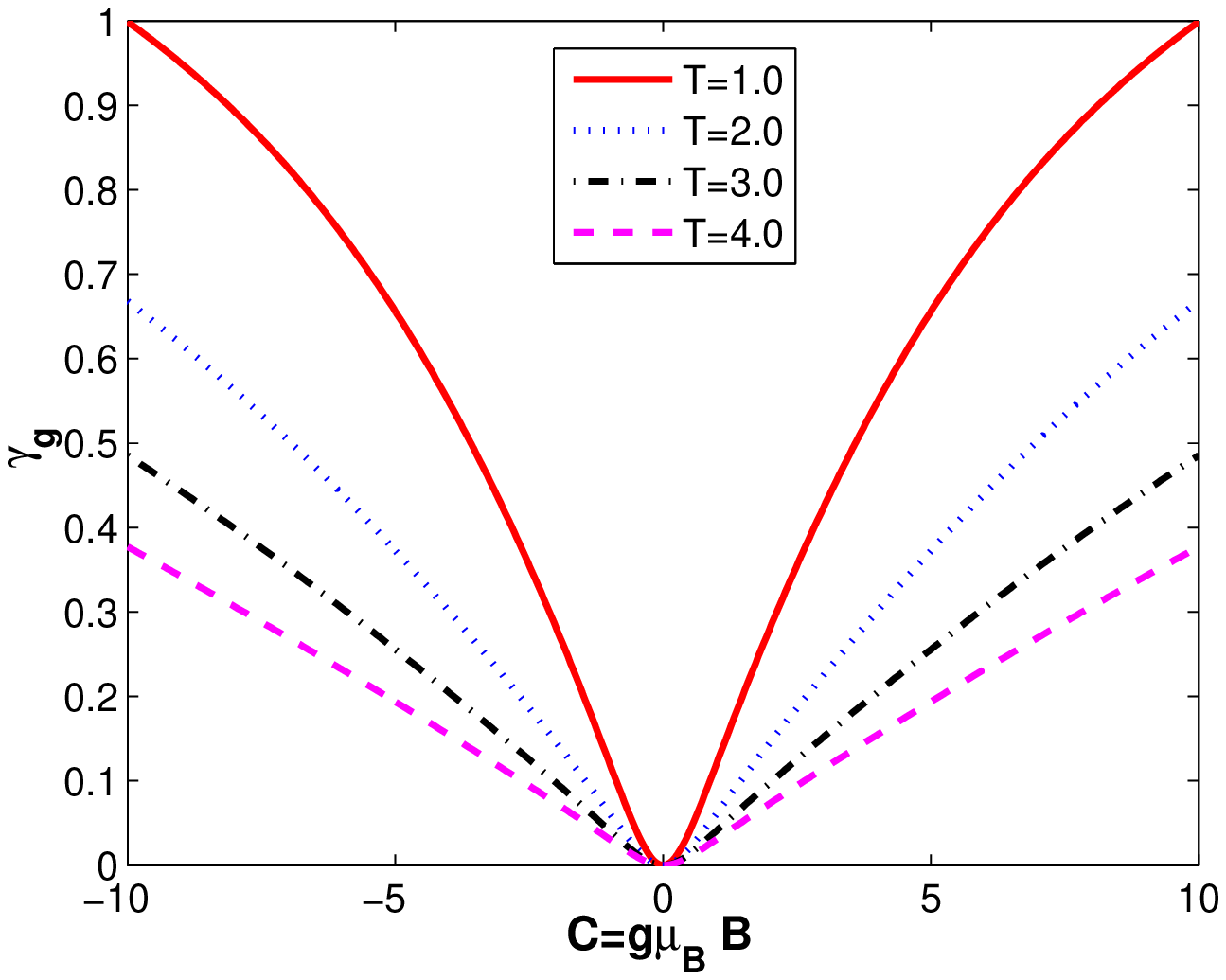}
\caption{\label{2order:C}Second order off-diagonal GP $\gamma_g$
versus external magnetic field $C=g\mu_B B$, when the temperature
$T=1.0,2.0,3.0,4.0 $ respectively.}
\end{center}
\end{figure}
Fig.\ref{2order:T} shows the relationship between 2-order
off-diagonal GP and the temperature. The $2$-order off-diagonal GP
can show other information which the diagonal GP cannot exhaust.
Also one can see increasing temperature tends to suppress the
2-order off-diagonal GP.

One can test the existence of off-diagonal geometric phase of the
above model. If one would detect such a phase the system should be
at low temperature, due to rapid decay of geometric phase with
temperature.

One can evaluate the mixedness of the state. The mixedness is
defined as $V=1-tr\rho(t)^2$. According to $\rho(t)=e^{-iHt}\rho_0
e^{iHt}$, one can easily obtain the mixedness of the state $\rho(t)$
as
\begin{equation}\label{mixedness}
V=1-tr\rho(t)^2=1-tr\left[U(t)\rho_0 U^{\dagger}(t)U(t)\rho_0
U^{\dagger}(t)\right]=\frac{6 (1+e^{\frac{J}{k
T}})}{(3+e^{\frac{J}{k
   T}})^2}.
\end{equation}
From Eq.\ref{mixedness}, one can see when the temperature
increasing, the mixedness is increased. When the temperature
approaches infinity, the mixedness approaches $3/4$. Note the
mixedness is not varied with the time, but it is the function of
temperature. With the increasing temperature of environment, the
thermal fluctuation of the system is enhanced and the coupling of
the system and environment thermal bath is also enhanced. The
enhanced thermal fluctuations suppress the GP of the system.
\section{Summary}
We have demonstrated the effect of temperature upon the off-diagonal
geometric phase. It shows that increasing temperature tends to
suppress the off-diagonal GP. Comparing ordinary diagonal GP,
off-diagonal GP can reveal more information which is not exhausted
by standard diagonal GP. As the temperature is increased, the
thermal fluctuation will dominate the behavior of the particles and
it will suppress the GP.

The work was supported by CJLU Grant No. 01101-000174.



\end{document}